\documentclass[11pt]{article}
\usepackage{amssymb,epsfig}
\begin{document}
\title{
\begin{flushright}
\normalsize
\begin{tabular}{r}
DFTT 24/98\\
UWThPh-1998-23\\
hep-ph/9805387
\end{tabular}
\end{flushright}
\vspace{1cm}
\textbf{Four-neutrino mixing and long-baseline neutrino oscillation
experiments}\thanks{Talk presented by C. Giunti at the
XXXIII$^{\mathrm{nd}}$
\textit{Rencontres de Moriond:
Electroweak Interactions and Unified Theories}, 
Les Arcs 1800 (France), March 14-21 1998.}
}
\author{
S.M. Bilenky\\
\textit{Joint Institute for Nuclear Research, Dubna, Russia}\\[3mm]
C. Giunti\\
\textit{INFN, Sezione di Torino, and Dipartimento di Fisica Teorica,
Universit\`a di Torino,}\\
\textit{Via P. Giuria 1, I--10125 Torino, Italy}\\[3mm]
W. Grimus\\
\textit{Institute for Theoretical Physics, University of Vienna,}\\
\textit{Boltzmanngasse 5, A--1090 Vienna, Austria}
}
\date{May 1998}
\maketitle
\begin{abstract}
We consider
the two schemes with four massive neutrinos
which are compatible with
the results of all neutrino oscillation experiments.
We show that in these two schemes
the probabilities of
$\bar\nu_e$
disappearance and
$
\stackrel{\makebox[0pt][l]
{$\hskip-3pt\scriptscriptstyle(-)$}}{\nu_{\mu}}
\to\stackrel{\makebox[0pt][l]
{$\hskip-3pt\scriptscriptstyle(-)$}}{\nu_{e}}
$
appearance in long-baseline experiments
are strongly suppressed.
\end{abstract}

\null \vspace{5mm}

The investigation of neutrino properties,
in particular of
neutrino masses and mixing,
is one of the most important problems
in today's high-energy physics and
many experiments are dedicated to it.
Among the numerous existing experimental results
there are three positive indications which come from
neutrino oscillation experiments.
Neutrino oscillations
can occur only if neutrinos are massive particles,
if their masses are different
and if neutrino mixing is realized in nature~\cite{BP78}.
In this case,
the left-handed flavor neutrino
fields
$\nu_{{\alpha}L}$
($\alpha=e,\mu,\tau$)
are superpositions
of
the left-handed components
$\nu_{kL}$
($k=1,\ldots,n$)
of the fields of neutrinos with definite masses
$m_k$:
\begin{equation}
\nu_{{\alpha}L}
=
\sum_{k=1}^{n}
U_{{\alpha}k}
\,
\nu_{kL}
\,,
\label{001}
\end{equation}
where $U$
is a unitary mixing matrix.
The general expression for the probability of
$\nu_\alpha\to\nu_\beta$
transitions in vacuum is
\begin{equation}
P_{\nu_\alpha\to\nu_\beta}
=
\left|
\sum_{k=1}^{n}
U_{{\beta}k}
\,
\exp\left( - i \, \frac{ \Delta{m}^2_{k1} \, L }{ 2 \, E } \right)
\,
U_{{\alpha}k}^*
\right|^2
\,,
\label{002}
\end{equation}
where
$\Delta{m}^2_{k1}$
is the mass-squared difference
$ m_k^2 - m_1^2 $,
$L$ is the distance between the neutrino source and detector
and $E$ is the neutrino energy.

The three experimental indications
in favor of neutrino oscillations come from the
solar neutrino problem~\cite{sunexp},
the atmospheric neutrino anomaly~\cite{atmexp}
and the results of the LSND experiment~\cite{LSND}.
The solar neutrino deficit can be explained
by transitions of $\nu_e$'s
into other states
due to a mass-squared difference
$ \Delta{m}^2_{\mathrm{sun}} \sim 10^{-5} \, \mathrm{eV}^2 $
in the case of resonant MSW transitions
or
$ \Delta{m}^2_{\mathrm{sun}} \sim 10^{-10} \, \mathrm{eV}^2 $
in the case of vacuum oscillations.
The atmospheric neutrino anomaly
can be explained
by transitions of $\nu_\mu$'s
into other states
due to a mass-squared difference
$ \Delta{m}^2_{\mathrm{atm}} \sim 5 \times 10^{-3} \, \mathrm{eV}^2 $.
The recent results of the first
reactor long-baseline (LBL)
neutrino oscillation experiment CHOOZ~\cite{CHOOZ}
exclude an explanation of the
atmospheric neutrino anomaly
through
$\nu_\mu\leftrightarrows\nu_e$
oscillations,
leaving the two possibilities:
$\nu_\mu\to\nu_\tau$
or
$\nu_\mu\to\nu_s$,
where $\nu_s$ is a sterile neutrino.
Finally,
the LSND experiment found
indications in favor of
$ \bar\nu_\mu \to \bar\nu_e $
oscillations
with a mass-squared difference in the range
$
0.3 \, \mathrm{eV}^2
\lesssim
\Delta{m}^2_{\mathrm{LSND}}
\lesssim
2.2 \, \mathrm{eV}^2
$,
which takes into account
the negative results of other
short-baseline (SBL) experiments~\cite{Brunner}.
Therefore,
the three experimental indications
in favor of neutrino oscillations
imply that there are
three independent mass-squared differences
and the number of massive neutrinos
is bigger than three.
In the following we consider the simplest possibility
of existence of four massive neutrinos.
In this case,
besides the three light flavor neutrinos
$\nu_e$,
$\nu_\mu$,
$\nu_\tau$
that
contribute to the invisible width of the $Z$-boson
measured with high accuracy by LEP experiments,
there is a light sterile flavor neutrino
$\nu_s$
that is a SU$(2)_L$ singlet
and does not take part in
the standard weak interactions.

Almost two years ago we have shown~\cite{BGG96}
that among all the possible
four-neutrino mass spectra
only two~\cite{OY96} are compatible
with the results of all neutrino oscillation experiments:
\begin{equation}
(\mathrm{A})
\quad
\underbrace{
\overbrace{m_1 < m_2}^{\mathrm{atm}}
\ll
\overbrace{m_3 < m_4}^{\mathrm{sun}}
}_{\mathrm{LSND}}
\,,
\qquad
(\mathrm{B})
\quad
\underbrace{
\overbrace{m_1 < m_2}^{\mathrm{sun}}
\ll
\overbrace{m_3 < m_4}^{\mathrm{atm}}
}_{\mathrm{LSND}}
\,.
\label{AB}
\end{equation}
In these two schemes
the four neutrino masses
are divided in two pairs of close masses
separated by a gap of about 1 eV,
which provides the mass-squared difference
$ \Delta{m}^2 \equiv \Delta{m}^2_{41} \equiv m_4^2 - m_1^2 $
that is relevant for the oscillations
observed in the LSND experiment.
In scheme A,
$ \Delta{m}^{2}_{21} \equiv m_2^2 - m_1^2 $
is relevant
for the explanation of the atmospheric neutrino anomaly
and
$ \Delta{m}^{2}_{43} \equiv m_4^2 - m_3^2 $
is relevant
for the suppression of solar $\nu_e$'s.
In scheme B,
the roles of
$\Delta{m}^{2}_{21}$
and
$\Delta{m}^{2}_{43}$
are reversed.

Let us define the quantities $c_\alpha$,
with $\alpha=e,\mu$,
in the schemes A and B as
\begin{equation}
(\mathrm{A})
\quad
c_\alpha
\equiv
\sum_{k=1.2}
|U_{{\alpha}k}|^2
\,,
\qquad
(\mathrm{B})
\quad
c_\alpha
\equiv
\sum_{k=3,4}
|U_{{\alpha}k}|^2
\,.
\label{04}
\end{equation}
The results of all neutrino oscillation experiments
are compatible with the schemes A and B
if~\cite{BGG96}
\begin{equation}
c_e \leq a_e^0
\quad \mbox{and} \quad
c_\mu \geq 1 - a_\mu^0
\,,
\label{03}
\end{equation}
where 
\begin{equation}
a_\alpha^0
\equiv
\frac{1}{2}
\left( 1 - \sqrt{ 1 - B_{\alpha;\alpha}^0 } \, \right)
\qquad
(\alpha=e,\mu)
\label{a0}
\end{equation}
and
$B_{\alpha;\alpha}^0$
is the upper bound for the amplitude of
$
\stackrel{\scriptscriptstyle(-)}{\nu}_{\hskip-3pt\alpha} 
\to
\stackrel{\scriptscriptstyle(-)}{\nu}_{\hskip-3pt\alpha}
$
oscillations
obtained from the exclusion plots of
SBL reactor and accelerator disappearance experiments.
Therefore,
the quantities 
$a_e^0$ and $a_\mu^0$
depend on $\Delta{m}^2$.
The exclusion curves obtained in the Bugey reactor experiment
and in the CDHS and CCFR accelerator
experiments~\cite{Bugey95-CDHS84-CCFR84}
imply that both
$a_e^0$ and $a_\mu^0$
are small quantities~\cite{BBGK96}:
$ a^{0}_e \lesssim 4 \times 10^{-2} $
and
$ a^{0}_\mu \lesssim 2 \times 10^{-1} $
for any value of
$\Delta{m}^{2}$
in the range
$
0.3
\lesssim
\Delta{m}^2
\lesssim
10^{3} \, \mathrm{eV}^2
$.

The smallness of $c_e$
in both schemes A and B
is a consequence of the solar neutrino problem~\cite{BGG96}.
It implies that the electron neutrino has a
small mixing with the neutrinos whose mass-squared difference is
responsible for the oscillations of atmospheric neutrinos
(\textit{i.e.},
$\nu_1$, $\nu_2$ in scheme A and $\nu_3$, $\nu_4$ in scheme
B).
Hence,
the transition probability of
electron neutrinos and antineutrinos
into other states
in atmospheric and long-baseline experiments
is suppressed.
Indeed,
it can be shown~\cite{BGG97}
that the transition probabilities
of electron neutrinos and antineutrinos
into all other states are bounded by
\begin{equation}
1 -
P^{(\mathrm{LBL})}_{\stackrel{\makebox[0pt][l]
{$\hskip-3pt\scriptscriptstyle(-)$}}{\nu_{e}}
\to\stackrel{\makebox[0pt][l]
{$\hskip-3pt\scriptscriptstyle(-)$}}{\nu_{e}}}
\leq
a^{0}_{e}
\left( 2 - a^{0}_{e} \right)
\,.
\label{05}
\end{equation}
The curve corresponding
to this limit
obtained from the 90\% CL exclusion plot of the Bugey
experiment is shown
in Fig.\ref{fig1}
(solid line).
The shadowed region in Fig.\ref{fig1}
corresponds to the range of $\Delta{m}^2$
allowed at 90\% CL by the results of the LSND
and all the other SBL experiments.
The upper bound for the transition
probability of $\bar\nu_e$'s into other states
obtained from the recently published
exclusion plot of the CHOOZ~\cite{CHOOZ} experiment
is shown in Fig.\ref{fig1} (dash-dotted line).
One can see that the CHOOZ upper bound
agrees with that obtained from Eq.(\ref{05}).
In Fig.\ref{fig1} we have also drawn the curve corresponding
to the expected final sensitivity of the CHOOZ experiment
(dash-dot-dotted line).
One can see that
the LSND-allowed region imply an upper bound for
$1-P^{(\mathrm{LBL})}_{\bar \nu_e \to \bar \nu_e}$
of about
$ 5 \times 10^{-2} $,
which is smaller than the expected final sensitivity of
the CHOOZ experiment.

\begin{table}[t!]
\begin{tabular*}{\textwidth}{@{\extracolsep{\fill}}cc}
\begin{minipage}{0.49\linewidth}
\begin{center}
\mbox{\epsfig{file=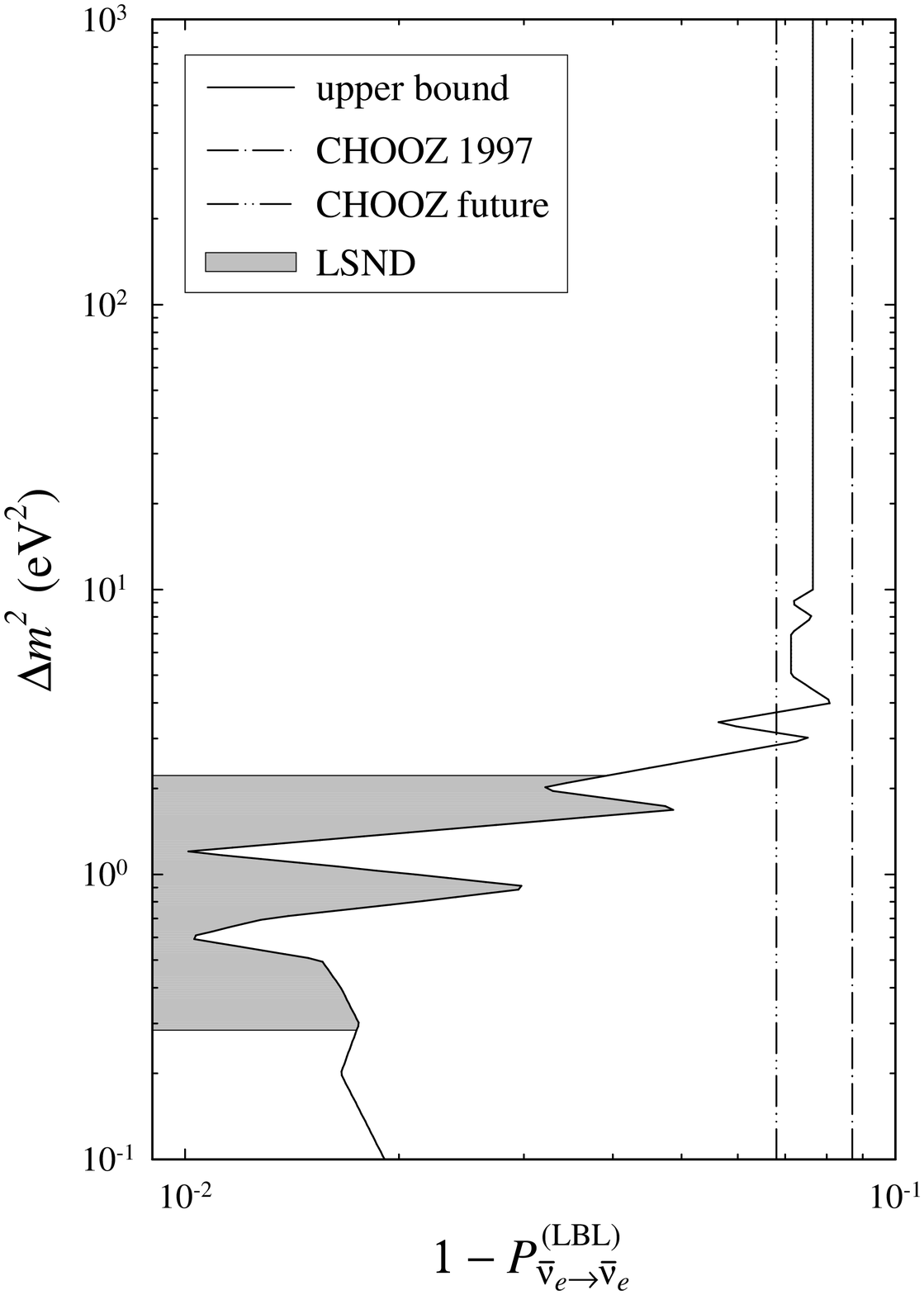,width=0.95\linewidth}}
\end{center}
\end{minipage}
&
\begin{minipage}{0.49\linewidth}
\begin{center}
\mbox{\epsfig{file=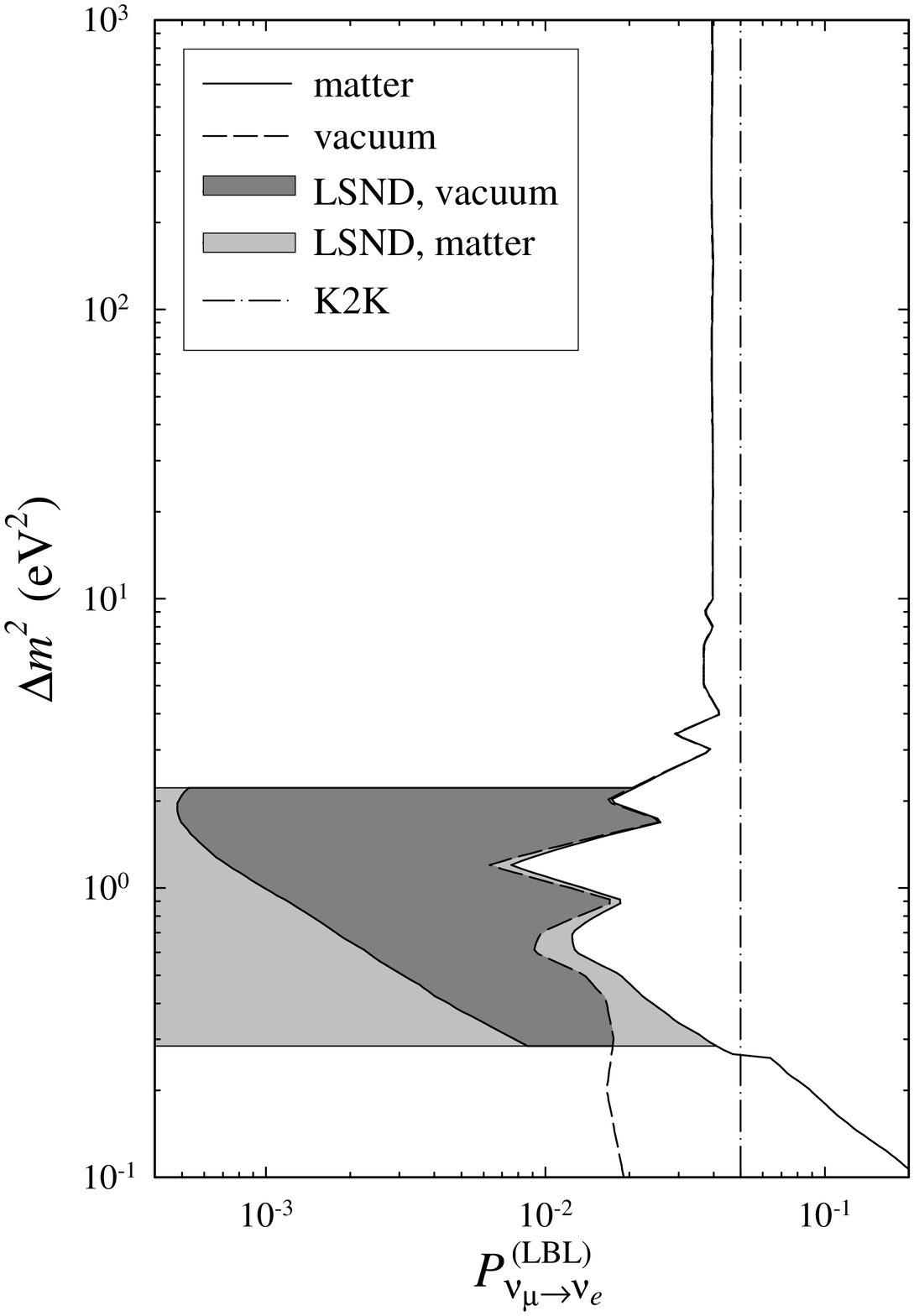,width=0.95\linewidth}}
\end{center}
\end{minipage}
\\
\refstepcounter{figure}
\label{fig1}                 
Figure \ref{fig1}
&
\refstepcounter{figure}
\label{fig2}                 
Figure \ref{fig2}
\end{tabular*}
\null \vspace{-0.5cm} \null
\end{table}

The probability of
$
\stackrel{\makebox[0pt][l]
{$\hskip-3pt\scriptscriptstyle(-)$}}{\nu_{\mu}}
\to\stackrel{\makebox[0pt][l]
{$\hskip-3pt\scriptscriptstyle(-)$}}{\nu_{e}}
$
transitions in LBL experiments
is bounded by~\cite{BGG97}
\begin{equation}
P^{(\mathrm{LBL})}_{\stackrel{\makebox[0pt][l]
{$\hskip-3pt\scriptscriptstyle(-)$}}{\nu_{\mu}}
\to\stackrel{\makebox[0pt][l]
{$\hskip-3pt\scriptscriptstyle(-)$}}{\nu_{e}}}
\leq
\min\!\left(
a^{0}_{e}
\left( 2 - a^{0}_{e} \right)
\, , \,
a^{0}_{e}
+
\frac{1}{4}
\,
A^{0}_{\mu;e}
\right)
,
\label{06}
\end{equation}
where
$A^{0}_{\mu;e}$
is the upper bound for the amplitude of
$
\stackrel{\makebox[0pt][l]
{$\hskip-3pt\scriptscriptstyle(-)$}}{\nu_{\mu}}
\to\stackrel{\makebox[0pt][l]
{$\hskip-3pt\scriptscriptstyle(-)$}}{\nu_{e}}
$
transitions measured in SBL experiments with accelerator neutrinos.
The bound obtained with Eq.(\ref{06})
from the 90\% CL exclusion plots of the Bugey
experiment
and
of the
BNL E734, BNL E776 and CCFR
experiments~\cite{BNLE734-BNLE776-CCFR96}
is depicted by the dashed line
in Fig.\ref{fig2}.
The solid line in Fig.\ref{fig2} shows the upper bound on
$
P^{(\mathrm{LBL})}_{\stackrel{\makebox[0pt][l]
{$\hskip-3pt\scriptscriptstyle(-)$}}{\nu_{\mu}}
\to\stackrel{\makebox[0pt][l]
{$\hskip-3pt\scriptscriptstyle(-)$}}{\nu_{e}}}
$
taking into account matter effects.
The expected sensitivities
of the K2K long-baseline accelerator neutrino experiment~\cite{K2K}
is indicated in Fig.\ref{fig2}
by the dash-dotted vertical line.
The shadowed region in Fig.\ref{fig2}
corresponds to the range of $\Delta{m}^2$
allowed at 90\% CL by the results of the LSND
and all the other SBL experiments.
It can be seen that the results of SBL experiments
indicate an upper bound for
$
P^{(\mathrm{LBL})}_{\stackrel{\makebox[0pt][l]
{$\hskip-3pt\scriptscriptstyle(-)$}}{\nu_{\mu}}
\to\stackrel{\makebox[0pt][l]
{$\hskip-3pt\scriptscriptstyle(-)$}}{\nu_{e}}}
$
smaller than
$ 4 \times 10^{-2} $
and
smaller than the expected sensitivity of
the K2K experiment.

In conclusion,
we have shown that
in the four-neutrino schemes A and B
which are compatible with
the results of all neutrino oscillation experiments,
the probabilities of
$\bar\nu_e$
disappearance and
$
\stackrel{\makebox[0pt][l]
{$\hskip-3pt\scriptscriptstyle(-)$}}{\nu_{\mu}}
\to\stackrel{\makebox[0pt][l]
{$\hskip-3pt\scriptscriptstyle(-)$}}{\nu_{e}}
$
appearance in LBL experiments
are severely constrained
(on the other hand,
the channels
$
\stackrel{\makebox[0pt][l]
{$\hskip-3pt\scriptscriptstyle(-)$}}{\nu_{\mu}}
\to\stackrel{\makebox[0pt][l]
{$\hskip-3pt\scriptscriptstyle(-)$}}{\nu_{\mu}}
$
and
$
\stackrel{\makebox[0pt][l]
{$\hskip-3pt\scriptscriptstyle(-)$}}{\nu_{\mu}}
\to\stackrel{\makebox[0pt][l]
{$\hskip-3pt\scriptscriptstyle(-)$}}{\nu_{\tau}}
$
are not constrained at all).
The two schemes A and B have
identical implications for neutrino oscillation experiments,
but very different implications
for neutrinoless double-$\beta$ decay experiments
and for tritium $\beta$-decay experiments.
Indeed,
in scheme A
\begin{equation}
|\langle{m}\rangle|
\leq
m_4
\qquad \mbox{and} \qquad
m(^3\mathrm{H}) \simeq m_4
\,,
\label{21}
\end{equation}
whereas in scheme B
\begin{equation}
|\langle{m}\rangle|
\leq
a_e^0 m_4
\ll
m_4
\qquad \mbox{and} \qquad
m(^3\mathrm{H}) \ll m_4
\,,
\label{22}
\end{equation}
where
$
\langle{m}\rangle
=
\sum_{i=1}^4 U_{ei}^2 m_i
$
is the effective Majorana mass
that determines the matrix element of neutrinoless double-$\beta$ decay
and
$m(^3\mathrm{H})$
is the neutrino mass measured in
tritium $\beta$-decay
experiments.
Therefore,
in scheme B
$|\langle{m}\rangle|$
and
$m(^3\mathrm{H})$
are smaller than the expected sensitivity
of the next generation of
neutrinoless double-$\beta$ decay
and tritium $\beta$-decay experiments.
The observation of a positive signal in these experiments
would be an indication in favor of scheme A.

\end{document}